\documentclass[12pt,oneside, a4paper]{article}

\ifx\pdfoutput\undefined
\usepackage[dvips,bookmarks=false]{hyperref}	
\else
\usepackage{hyperref}	
\fi
\hypersetup{colorlinks,bookmarksopen,bookmarksnumbered,citecolor=blue,
linkcolor=black,pdfstartview=FitH,urlcolor=blue}

\newcommand{\1}{\mbox{1}\hspace{-0.25em}\mbox{l}}

\usepackage{graphicx}
\usepackage{amssymb}
\usepackage{cite}
\usepackage{bm}
\usepackage{indentfirst}
\usepackage{amsmath}
\usepackage{hhline}
\usepackage{multirow}
\usepackage{here}

\newcommand{\bhline}[1]{\noalign{\hrule height #1}}

\begin{document}

\begin{titlepage}

\begin{flushright}
IFIC/15-41\\
KANAZAWA-15-11\\
LPT-Orsay-15-42
\end{flushright}

\begin{center}

\vspace{1cm}
{\large\bf 
Non-thermal Production of Minimal Dark Matter via Right-handed Neutrino Decay
}
\vspace{1cm}

\renewcommand{\thefootnote}{\fnsymbol{footnote}}
Mayumi Aoki$^1$\footnote[1]{mayumi@hep.s.kanazawa-u.ac.jp},
Takashi Toma$^2$\footnote[2]{takashi.toma@th.u-psud.fr},
Avelino Vicente$^{3,4}$\footnote[3]{Avelino.Vicente@ulg.ac.be} 
\vspace{5mm}

{\it%
$^1${Institute for Theoretical Physics, Kanazawa University, Kanazawa
 920-1192, Japan}\\
$^2${Laboratoire de Physique Th\'eorique, CNRS - UMR 8627,\\
 Universit\'e de Paris-Sud 11, F-91405 Orsay Cedex, France}\\
$^3${IFPA, Dep. AGO, Universit\'e de Li\`ege, Bat B5, Sart-Tilman B-4000
Li\`ege 1, Belgium}\\
$^4${Instituto de F\'{\i}sica Corpuscular (CSIC-Universitat de Val\`{e}ncia), \\
Apdo. 22085, E-46071 Valencia, Spain}
}

\vspace{8mm}

\abstract{ Minimal Dark Matter (MDM) stands as one of the simplest
  dark matter scenarios. In MDM models, annihilation and
  co-annihilation processes among the members of the MDM multiplet are
  usually very efficient, pushing the dark matter mass above
  $\mathcal{O}(10)$ TeV in order to reproduce the observed dark matter
  relic density. Motivated by this little drawback, in this paper we
  consider an extension of the MDM scenario by three right-handed
  neutrinos. Two specific choices for the MDM multiplet are studied: a
  fermionic $SU(2)_L$ quintuplet and a scalar $SU(2)_L$ septuplet. The
  lightest right-handed neutrino, with tiny Yukawa couplings, never
  reaches thermal equilibrium in the early universe and is produced by
  freeze-in. This creates a link between dark matter and neutrino
  physics: dark matter can be non-thermally produced by the decay of
  the lightest right-handed neutrino after freeze-out, allowing to
  lower significantly the dark matter mass. We discuss the
  phenomenology of the non-thermally produced MDM and, taking into
  account significant Sommerfeld corrections, we find that the dark
  matter mass must have some specific values in order not to be in
  conflict with the current bounds from gamma-ray observations. }

\end{center}
\end{titlepage}

\renewcommand{\thefootnote}{\arabic{footnote}}
\setcounter{footnote}{0}

\setcounter{page}{2}

\section{Introduction}
\label{sec:int}

The identity of the Dark Matter (DM) that makes up about $25 \%$ of
the energy content of the universe is one of the most important open
problems in (astro-)particle physics. Lots of candidates have been
proposed under the assumption that DM is made of particles. The most
popular options include Weakly Interacting Massive Particles (WIMPs),
axions, gravitinos and asymmetric dark matter. In particular, WIMPs
are theoretically well-motivated candidates since the present relic
density of DM can be reproduced by thermal production in the early
universe with an electroweak scale DM mass and an annihilation cross
section
$\langle\sigma{v}\rangle\approx3\times10^{-26}~\mathrm{cm^3/s}$, in
the typical range for a particle with weak interactions. This
intriguing coincidence, usually called \emph{the WIMP miracle}, has
triggered a massive exploration of WIMP DM scenarios, with detailed
studies of their phenomenological implications and dedicated
experimental searches for DM in the form of WIMPs in colliders as well
as in direct and indirect detection experiments.

In scenarios with electroweak scale DM, a discrete symmetry is often
imposed in order to stabilize the DM candidate. Although this symmetry
is usually introduced \emph{by hand}, many theoretical justifications
are known. For instance, this symmetry can be seen as a remnant after
the spontaneous breaking of a larger symmetry group. Many proposals in
this direction exist, based on global~\cite{Lindner:2011it,
Baek:2013fsa, Bernal:2015bla} or gauge
symmetries~\cite{Hambye:2008bq, Ko:2014nha, Baek:2014kna}, some of them
linked to flavor 
symmetries~\cite{Hirsch:2010ru,Sierra:2014kua}. A completely different
approach is to consider that the origin of this symmetry is
accidental. If a large multiplet of the $SU(2)_L$ gauge symmetry of
the Standard Model (SM) is introduced, an accidental $\mathbb{Z}_2$
symmetry may appear due to the restrictions imposed by gauge
invariance and renormalizability. This is the so-called \emph{Minimal
  Dark Matter} (MDM) approach~\cite{Cirelli:2005uq}, a popular
scenario with some recent works~\cite{Geng:2014oea, Yu:2015pwa,
  Culjak:2015qja, Harigaya:2015yaa, Ahriche:2015wha}.

There is, however, a generic drawback in MDM scenarios: the components
of the large $SU(2)_L$ multiplets are generally required to be nearly
degenerate. The origin of this mass degeneracy is easy to
understand. First of all, in some MDM models this is actually
predicted, since the mass splittings among components of the large
$SU(2)_L$ multiplets only appear at the one-loop
level~\cite{Cirelli:2009uv,Hambye:2009pw}. When this is not the case,
and large mass splittings can in principle be obtained, one must face
two problems. First, the large $SU(2)_L$ multiplets contribute to
electroweak precision observables (EWPO) through the $STU$ oblique
parameters and, in order to be compatible with the current
experimental measurements, one typically requires small mass
splittings~\cite{Cai:2012kt,Earl:2013jsa}. And second, the mass
splittings must be induced by Higgs-DM-DM couplings for scalar DM,
which are required to be small in order to suppress the elastic
scattering with nuclei via Higgs exchange and be compatible with
direct detection constraints (see for example
\cite{Cline:2013gha}). As a result of this degeneracy, all members of
the multiplet will be in thermal equilibrium during freeze-out,
co-annihilating very efficiently with the DM particles and strongly
reducing the DM relic density. In order to reproduce the relic density
measured by Planck, $\Omega_{\text{DM}} h^2=0.1186\pm0.0020$ at $68\%$
Confidence Level (CL)~\cite{Adam:2015rua}, this implies a heavy DM
particle.  In fact, once Sommerfeld corrections are
included~\cite{Cirelli:2007xd}, the DM mass is typically found to be
about $\mathcal{O}(10)$ TeV. Although perfectly plausible, this is not very
attractive from a phenomenological point of view.

Another open problem that calls for physics beyond the SM is the
existence of non-zero neutrino masses and mixings, as established by
neutrino oscillation experiments. Many extensions of the SM have been
proposed to address this issue. The energy scale for the new states
responsible for neutrino mass generation can be either very high,
potentially relating neutrino masses to unification, or low (TeV scale
or below). In the latter case, the existence of new physics at low
energies leads to interesting phenomenological perspectives, within
the reach of current collider and low-energy experiments. Furthermore,
many neutrino mass models include DM candidates, although an interplay
between these two fundamental issues is often missing.

In this paper, we consider a very economical extension of the MDM
scenario. In addition to the multiplet containing the DM particle,
three right-handed neutrino singlets are introduced.  No additional
symmetry for DM stabilization is required due to an accidental
$\mathbb{Z}_2$ symmetry resulting from the gauge invariant
renormalizable interactions of the DM multiplet. The right-handed
neutrinos play two roles in this model. First, neutrino masses are
induced at tree-level with the standard Type-I Seesaw
mechanism~\cite{Minkowski:1977sc,seesaw,Mohapatra:1979ia,Schechter:1980gr,Schechter:1981cv},
and second, the out-of-equilibrium decay of the right-handed neutrinos
leads to non-thermal production of DM, allowing one to compensate the
strong effect of co-annihilations and lower the DM mass below the TeV
scale.\footnote{Non-thermal production of Wino DM (triplet) has been discussed in
Ref.~\cite{Moroi:2013sla}.} This setup will be illustrated with two specific
examples: a model with a fermionic quintuplet and a model with a scalar septuplet, in
both cases with vanishing hypercharge.\footnote{Septuplets with
hypercharge $Y=2$ have also been considered in Refs.~\cite{Hisano:2013sn,
Alvarado:2014jva}, although with a completely different
motivation: the extended model keeps the $\rho$-parameter as $1$ at
tree-level.} The number of physical parameters in both models is
limited and many experimental constraints exist. 
As a result of this, the setup is very predictive
and will definitely be tested in future experiments.

The rest of the paper is organized as follows: in Sec.~\ref{sec:model}
we introduce the setup, comment on the degeneracy among the members of
the MDM multiplet and discuss the mechanism responsible for neutrino
mass generation. The DM phenomenology of these models is explored in
Sec.~\ref{sec:DMpheno}, where we present the main results of this
paper. The most relevant constraints in our scenario are discussed in
Sec.~\ref{sec:constraints} and, finally, we summarize and present the
main conclusions of the paper in Sec.~\ref{sec:sum}.

\section{The Models}
\label{sec:model}

\subsection{New particles and interactions}

\begin{table}[t]
\centering
\begin{tabular}{cccc}\bhline{1pt}
          && $N_i$    & $\chi$ \\\bhline{1pt}
$SU(2)_L$ && ${\bf 1}$ & ${\bf 5}$ (${\bf 7}$)\\\hline
$U(1)_Y$  && $0$       & $0$      \\\hline
spin  && $1/2$       & $1/2$ ($0$)      \\
\bhline{1pt}
\end{tabular}
\caption{New particle content and charge assignment in the two models
  under consideration. Here $i=1,2,3$.  The first model introduces a
  quintuplet fermion, whereas the second introduces a septuplet
  scalar, in both cases with $Y=0$.  }
\label{tab:content}
\end{table}

We consider an extension of the SM by three right-handed neutrinos
$N_i~(i=1-3)$ and a $SU(2)_L$ multiplet $\chi$ which is either a quintuplet
fermion or a septuplet scalar, as summarized in
Tab.~\ref{tab:content}. The quintuplet fermion and the septuplet
scalar can be denoted by the vectors
\begin{equation}
\chi\equiv i\left(
\begin{array}{c}
+\chi^{++}\\
+\chi^{+}\\
-\chi^0\\
+\chi^{-}\\
+\chi^{--}
\end{array}
\right),\quad 
\chi\equiv i\left(
\begin{array}{c}
+\chi^{+++}\\
+\chi^{++}\\
+\chi^{+}\\
-\chi^0\\
-\chi^{-}\\
+\chi^{--}\\
-\chi^{---}
\end{array}
\right) \, .
\label{eq:chi}
\end{equation}
The prefactor $i$ and the sign for each component are taken in order to
satisfy $\chi^c=\chi$ where $\chi^c$ denote charge conjugation of the
field $\chi$.
Further details about products of $SU(2)$ multiplets are given in Appendix \ref{sec:app}. 
The kinetic and Yukawa Lagrangians of the new
particles are given by
\begin{equation}
\mathcal{L} =
\mathcal{L}_K^{\mathrm{MDM}}
+\frac{1}{2}\overline{N_i^c}\left(i\partial\!\!\!/-m_{N_i}\right)N_i
-\left(y_{i\alpha}^\nu\phi\overline{N_i}P_LL_\alpha+\mathrm{H.c.}\right) \, ,
\label{eq:lag1}
\end{equation}
where $\mathcal{L}_K^{\mathrm{MDM}}$ is the kinetic term of the multiplet given by 
\begin{equation}
\mathcal{L}_K^{\mathrm{MDM}}=\left\{
\begin{array}{cc}
\displaystyle
\frac{1}{2}\overline{\chi^c}\left(iD\!\!\!\!/-m_\chi\right)\chi
\vspace{0.2cm}
&\quad\mbox{for the quintuplet fermion,}\\
\displaystyle
\frac{1}{2}\left(D_\mu\chi\right)\left(D^\mu\chi\right)
&\quad\mbox{for the septuplet scalar.}
\end{array}
\right.
\label{eq:lag2}
\end{equation}
The covariant derivative is defined by $D_\mu \equiv \partial_\mu + ig_2
W_\mu^a T^a$ with the SU(2)$_L$ gauge coupling $g_2$, gauge field
$W_\mu^a$ and generator $T^a$ $(a=1,2,3)$.
Note that Eq.~\eqref{eq:lag1} is written in the basis in
which the right-handed neutrino Majorana mass term is diagonal.
From the Lagrangian, one can see that the right-handed neutrinos
dominantly interact with the longitudinal mode of gauge bosons $W^\pm$
and $Z$, while the MDM multiplet $\chi$ interacts with the transverse
mode. 
The scalar potential for the model with the quintuplet fermion is
exactly the same as in the SM ($\mathcal{V}_5 =
\mathcal{V}_{\text{SM}}$), whereas that for the septuplet scalar model
is
\begin{equation}
\mathcal{V}_7
=
-\mu_\phi^2|\phi|^2
+\frac{\mu_\chi^2}{2} \chi^2
+\frac{\lambda_\phi}{4}|\phi|^4
+\sum_{k=1}^2\frac{\lambda_{\chi k}}{4!}\left[\chi^4\right]_k
+\frac{\lambda_{\phi\chi}}{2}|\phi|^2\chi^2 \, ,
\label{eq:potential}
\end{equation}
where $\phi$ is the SM Higgs doublet with hypercharge $Y_\phi = 1/2$. 
In the following, the scalar coupling $\lambda_{\phi\chi}$ is assumed to
be small, as in Ref.~\cite{Cirelli:2005uq}. 

Several comments about the scalar potential in
Eq.~\eqref{eq:potential} are in order. First, we note that in some
cases various singlet contractions of $SU(2)_L$ higher dimensional
representations are possible. For example, four kinds of singlets can
be constructed for the $\chi^4$ term. However only two of them are
independent. This is denoted with a summation in
Eq.~\eqref{eq:potential}.  Furthermore, there is no septuplet cubic
term $\chi^3$, since the singlet obtained by contracting three
septuplets is completely anti-symmetric:
${\bf7}\otimes{\bf7}\otimes{\bf7}\supset{\bf1}_A$.  As a result, an
accidental $\mathbb{Z}_2$ symmetry appears, under which the septuplet
$\chi$ is odd and the rest of the fields even.  The same accidental
symmetry appears for the quintuplet fermion as well.  Therefore, the
lightest state contained in the quintuplet or septuplet will be stable
and a DM candidate is automatically included in the model without any
additional symmetry.\footnote{For the septuplet scalar, we will assume
  that the vacuum of the theory is such that no other scalar field
  besides the usual Higgs doublet has a vacuum expectation value
  (VEV). This guarantees that the accidental $\mathbb{Z}_2$ symmetry
  remains after electroweak symmetry breaking.}  One should note that
for the septuplet scalar, the dimension $5$ operator
$\chi^3|\phi|^2/\Lambda$, where $\Lambda$ is the energy scale at which the
operator is induced, leads to the decay of the DM candidate at one-loop
level and is not forbidden by any symmetry. 
As it has been recently discussed in Ref.~\cite{DiLuzio:2015oha},
$\Lambda\gtrsim10^{20}~\mathrm{GeV}$ is necessary for
$m_\chi\sim10~\mathrm{TeV}$ in order to obtain a DM lifetime longer than the age
of the universe, $\tau_U\sim10^{18}~\mathrm{s}$. Moreover, pairs of gauge bosons 
$\gamma\gamma$, $\gamma Z$, $ZZ$ and $W^+W^-$ are
produced by the DM decays, and these are constrained by gamma-ray
experiments. One finds that the DM lifetime should be roughly
$\tau_\chi\gtrsim10^{27}~\mathrm{s}$ to evade them. 
This corresponds to $\Lambda\gtrsim10^{25}~\mathrm{GeV}$ which is much
larger than the Planck scale.
In contrast, there is no possible $5$
dimensional operator for the quintuplet fermion
case~\cite{DiLuzio:2015oha}. 
The possible $6$ dimensional operators are $\chi
L_\alpha|\phi|^2\phi/\Lambda^2$ and $\chi\sigma^{\mu\nu}L_\alpha\phi
W_{\mu\nu}/\Lambda^2$, where $W_{\mu\nu}$ is the field strength tensor for
the $SU(2)_L$ gauge group. For such $6$ dimensional operators, a
DM lifetime long enough to satisfy the gamma-ray constraints can be
achieved with $\Lambda=10^{15}~\mathrm{GeV}$. 
Thus, the quintuplet fermion DM candidate
would be stable even if one includes non-renormalizable operators.

Finally, lepton number is softly broken by the Majorana mass term of
the right-handed neutrinos. As a consequence, neutrinos acquire
Majorana masses through the canonical type-I seesaw mechanism, as we
will see below.

\subsection{Mass degeneracy within the MDM multiplets}

After electroweak symmetry breaking, only the SM Higgs doublet gets a
non-zero VEV, $\phi^0=\langle\phi\rangle+h/\sqrt{2}$.  The Higgs boson
mass is given by $m_h^2=\lambda_\phi\langle\phi\rangle^2$ and all the
components of the multiplet $\chi$ have the same mass $m_\chi$ at
tree-level. The mass $m_\chi$ is given by the bare mass in
Eq.~(\ref{eq:lag2}) for the quintuplet fermion and
$m_\chi^2=\mu_\chi^2+\lambda_{\phi\chi}\langle\phi\rangle^2$ for the
septuplet scalar.  However, a mass difference is induced at the one-loop
level. Given two components of the MDM multiplets with electric
charges $Q$ and $Q'$, the one-loop induced mass splitting is computed to
be
\begin{equation}
m_{Q}-m_{Q'}=\left(Q^2-Q'^2\right)
\frac{m_\chi}{4\pi}
\left[
\alpha_W\left\{
f\left(\frac{m_W}{m_\chi}\right)
-f\left(\frac{m_Z}{m_\chi}\right)
\right\}
+\alpha_{\mathrm{em}}\left\{
f\left(\frac{m_Z}{m_\chi}\right)
-f\left(0\right)
\right\}
\right], 
\label{eq:dmass}
\end{equation}
where $\alpha_W=g_2^2/(4\pi)$, $\alpha_\mathrm{em}=e^2/(4\pi)$ and the
function $f(z)$ is defined as
\begin{equation}
f\left(z\right)=
2\int_0^1\left(1+x\right)\log\Bigl((1-x)z^2+x^2\Bigr)dx,
\label{eq:loop_f1}
\end{equation}
for the quintuplet fermion and 
\begin{equation}
f\left(z\right)=
-\frac{1}{2}\int_0^1
\Bigl(6\left(1-x\right)z^2+9x^2-4x-4\Bigr)\log\Bigl(\left(1-x\right)z^2+x^2
\Bigr)
dx,
\label{eq:loop_f2}
\end{equation}
for the septuplet scalar.  The septuplet scalar not only has $SU(2)_L$
gauge interactions, but also scalar couplings $\lambda_{\chi k}$ and
$\lambda_{\phi\chi}$ in the potential. These couplings also give a
correction to the scalar masses. However, the mass corrections for all
the components of the septuplet are exactly the same and no mass
difference is generated in this way. We note that, although it may
seem that our results differ from those in Ref.~\cite{Cirelli:2005uq},
we have checked explicitly that our expressions in
Eqs.~(\ref{eq:dmass})-(\ref{eq:loop_f2}) are consistent with those in
this reference.

Inspection of Eqs.~(\ref{eq:dmass})-(\ref{eq:loop_f2}) reveals some
relevant features of the mass degeneracy within the MDM
multiplets. For light multiplets ($m_\chi \ll m_{W},m_Z$) the
corrections are negligible for quintuplets but can reach a few GeV in
the case of septuplets. Then, as the multiplet mass increases, the
splitting in both cases approaches a common value. In fact, in the
$m_\chi\to\infty$ limit one finds
\begin{equation} \label{eq:dmassapp}
m_Q-m_{Q'}=(Q^2-Q'^2) \, \alpha_W \, m_W \, \sin^2
 \left(\frac{\theta_W}{2}\right) 
\approx(Q^2-Q'^2)\times166~\mathrm{MeV},
\end{equation}
in both MDM scenarios (quintuplet and septuplet). Therefore, although
the behavior of the loop function $f(z)$ is different for
low $m_\chi$ values, it is the same for $m_\chi \gg m_{W},m_Z$, leading
to a \emph{universal} splitting in case of heavy MDM multiplets. This
universality can be used to estimate the resulting splitting at the
two-loop level. The mass splitting between the singly charged and
neutral components of a triplet fermion was calculated at the two-loop
level in Ref.~\cite{Ibe:2012sx}, finding the value $164.4~\mathrm{MeV}$ in
the limit of an infinitely heavy triplet. Given the universality of
this limit, we expect the same conclusion to hold in our two MDM
scenarios. Finally, we also note that the previous expressions imply
that the neutral component $\chi^0$, is the lightest of the
components of the MDM multiplet, thus becoming a viable DM candidate.

\subsection{Neutrino mass matrix}

The Lagrangian relevant for the generation of neutrino masses is
\begin{equation}
\mathcal{L}_\nu=-\left(m_D\right)_{i\alpha}\overline{N_i}P_L\nu_\alpha
-\frac{m_{N_i}}{2}\overline{N_i^c} N_i+\mathrm{H.c.}, 
\end{equation}
where $\left(m_D\right)_{i\alpha}=y^\nu_{i\alpha}\langle\phi\rangle$ is
the usual neutrino Dirac mass term. Assuming
$\left(m_D\right)_{i\alpha} \ll m_{N_i}$, the light neutrinos acquire
Majorana masses via the canonical type-I seesaw mechanism,
\begin{equation}
m_\nu \approx -m_D^T m_N^{-1} m_D \, .
\label{eq:nu-mass}
\end{equation}
The light neutrino mass matrix in Eq.~(\ref{eq:nu-mass}) is
diagonalized by the Pontecorvo-Maki-Nakagawa-Sakata matrix $U_{\mathrm{PMNS}}$
as
\begin{equation}
U_\mathrm{PMNS}^T \, m_\nu \, U_\mathrm{PMNS}=\left(
\begin{array}{ccc}
m_1 & 0 & 0\\
0 & m_2 & 0\\
0 & 0 & m_3
\end{array}
\right) \, .
\end{equation}
It is common to use the Casas-Ibarra
parametrization to write the neutrino Yukawa
coupling $y^\nu$ as~\cite{Casas:2001sr}
\begin{equation}
y^\nu = \frac{1}{\langle\phi\rangle}D_{\sqrt{m_N}} R D_{\sqrt{m}} U_{\mathrm{PMNS}}^\dagger \, , 
\end{equation}
where $D_{\sqrt{m_N}} = \text{diag}\left( \sqrt{m_{N_i}} \right)$,
$D_{\sqrt{m}} = \text{diag}\left( \sqrt{m_i} \right)$, $R$ is an
orthogonal matrix ($R^T R = R R^T = \1$) and the extra phase of the
Yukawa coupling has been absorbed into the fields.

As we will see below, non-thermal production of DM requires
right-handed neutrinos with masses
$m_{N_i}\gtrsim5~\mathrm{TeV}$. Therefore, the Yukawa couplings of the
right-handed neutrinos participating in the generation of the light
neutrino masses will be of the order of $y^\nu\gtrsim10^{-6}$ in order
to obtain an appropriate neutrino mass scale.

\section{Non-thermal production of Minimal Dark Matter}
\label{sec:DMpheno}

The most recent measurement of the DM relic density by the Planck
Collaboration is $\Omega_{\text{DM}} h^2=0.1186\pm0.0020$ at $68\%$
CL~\cite{Adam:2015rua}. In this section we will study the implications
of this measurement for our setup, where the lightest neutral
component of the MDM multiplet $\chi^0$, is the DM candidate due to
the accidental $\mathbb{Z}_2$ symmetry.

A DM particle with $SU(2)_L$ gauge interactions and a mass larger than
the $W$ boson mass would mainly annihilate into a pair of gauge
bosons, typically implying a strong reduction of the relic
density. Thus, in the case of the MDM scenario with a large
dimensional representation~\cite{Cirelli:2005uq}, the DM mass has to
be above several TeV in order to reproduce the observed relic
density. Moreover, the components of the MDM multiplet are always
required to be nearly degenerate. This enhances the co-annihilation
cross sections and also leads the Sommerfeld correction to the cross
sections. 
Therefore this implies even heavier DM particles, with masses
above $10~\mathrm{TeV}$~\cite{Hisano:2006nn,Cirelli:2007xd}. Even if
the DM mass is below the $W$ boson mass, efficient co-annihilation with the
charged particles in the MDM multiplet is still at
work, and the DM relic density would be too low.

In the absence of the heavy neutrinos, the above problem would be
present in our models. However, the picture is slightly changed due to
the non-thermal production of DM in the out-of-equilibrium decay of
the right-handed neutrinos. The lightest right-handed neutrino is
expected to be produced by a freeze-in mechanism in the early universe
if its Yukawa couplings are small
enough~\cite{McDonald:2001vt,Hall:2009bx,Blennow:2013jba,Klasen:2013ypa,
Molinaro:2014lfa}.
Then, once produced, it mainly decays into two body final states,
$N_1\to h\nu_\alpha,Z\nu_\alpha,W^\pm\ell_\alpha^\mp$. The decay
width into two body final states is computed as
\begin{equation}
\Gamma_{N_1}=\frac{\left(y^\nu y^{\nu\dag}\right)_{11}m_{N_1}}{8\pi},
\label{eq:width}
\end{equation}
where the masses of the gauge and Higgs bosons have been
neglected.\footnote{Note that in order to simplify the notation we
  have decided to denote the $N_1$ decay width into two body final
  states as $\Gamma_{N_1}$. However, this should not be confused with
  the $N_1$ total decay width, which would also include three body
  final states.}  There are also subdominant three body $N_1$ decay
processes into the components of the multiplet, such as
$N_1\to\chi^\pm\chi^\mp\nu_\alpha$ and
$N_1\to\chi^0\chi^\pm\ell^\mp_\alpha$, mediated by the gauge bosons.
Then the charged particles decay into $\chi^0$.  These three body
decay processes occur due to the mixing between left-handed and
right-handed neutrinos.  The branching ratio of these processes,
including all the components of the MDM multiplet, will be denoted as
$\mathrm{Br}_\chi$. The number of DM particles produced per $N_1$
decay is $2 \, \mathrm{Br}_\chi$, since a pair of DM particles is
produced in each decay of $N_1$ due to the conservation of the
accidental $\mathbb{Z}_2$ parity. In our analysis we will consider
$\mathrm{Br}_\chi$ as a free parameter. Although the minimal models
discussed in this paper predict a too low $\mathrm{Br}_\chi$ value, we
will comment below on how to increase this parameter with a minimal
extension.

\begin{figure}[t]
\begin{center}
\includegraphics[scale=0.65]{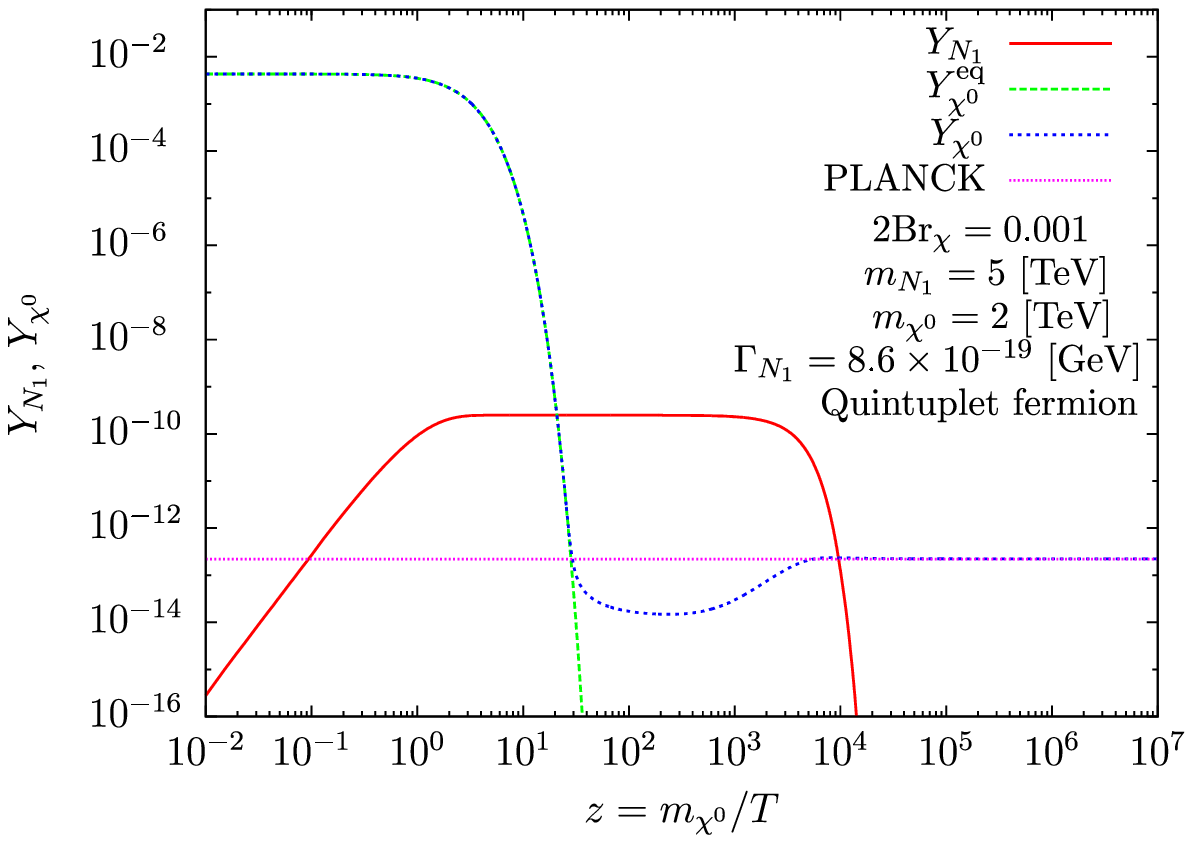}
\includegraphics[scale=0.65]{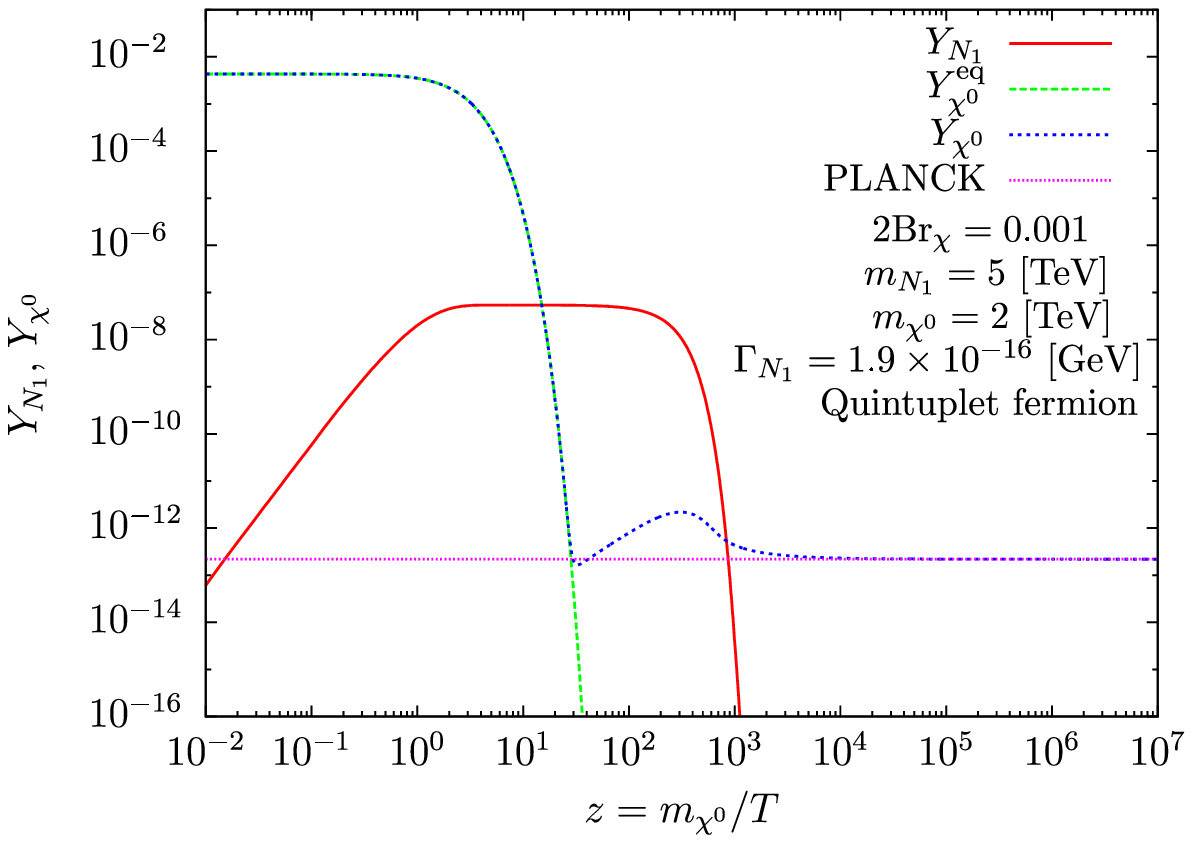}
\caption{Examples of solutions of the coupled Boltzmann equations.}
\label{fig:boltzmann}
\end{center}
\end{figure}

The evolution of the number densities of the $N_1$ and $\chi^0$
species in the early universe is given by the coupled Boltzmann
equations
\begin{eqnarray}
\frac{dn_{N_1}}{dt}+3Hn_{N_1}\hspace{-0.2cm}&=&\hspace{-0.2cm}
\frac{g_{N_1}m_{N_1}^2m_{\chi^0}\Gamma_{N_1}}{2\pi^2
z}K_1\left(\frac{m_{N_1}}{m_{\chi^0}}z\right)
-\Gamma_{N_1}n_{N_1},\\
\frac{dn_{\chi^0}}{dt}+3Hn_{\chi^0}\hspace{-0.2cm}&=&\hspace{-0.2cm}
-\langle\sigma_\mathrm{eff}{v}\rangle\left(n_{\chi^0}^2-{n_{\chi^0}^\mathrm{eq}}^2\right)
+2 \, \mathrm{Br}_\chi\Gamma_{N_1}n_{N_1} \, .
\end{eqnarray}
Here $z=m_{\chi^0}/T$ with the temperature of the universe $T$,
$g_{N_1}=2$ is the number of degrees of freedom 
of $N_1$, $K_1(z)$ is the second modified Bessel function,
$n_{\chi^0}^\mathrm{eq}$ is the equilibrium number density of $\chi^0$
and $\langle\sigma_\mathrm{eff}{v}\rangle$ is the thermally averaged
effective DM annihilation cross section, including co-annihilation
processes with degenerate particles.  The dominant annihilation and
co-annihilation channels are $\chi^0\chi^0\to W^\pm W^\mp$ and
$\chi^0\chi^\pm,\chi^\mp\chi^{\pm\pm}~(\chi^{\mp\mp}\chi^{\pm\pm\pm})\to
W^{\pm*}\to\gamma W^\pm$.  In our analysis we approximately include
the Sommerfeld effect for the effective annihilation cross section by
using the results obtained in Ref.~\cite{Cirelli:2007xd}.

\begin{figure}[t]
\begin{center}
\includegraphics[scale=0.57]{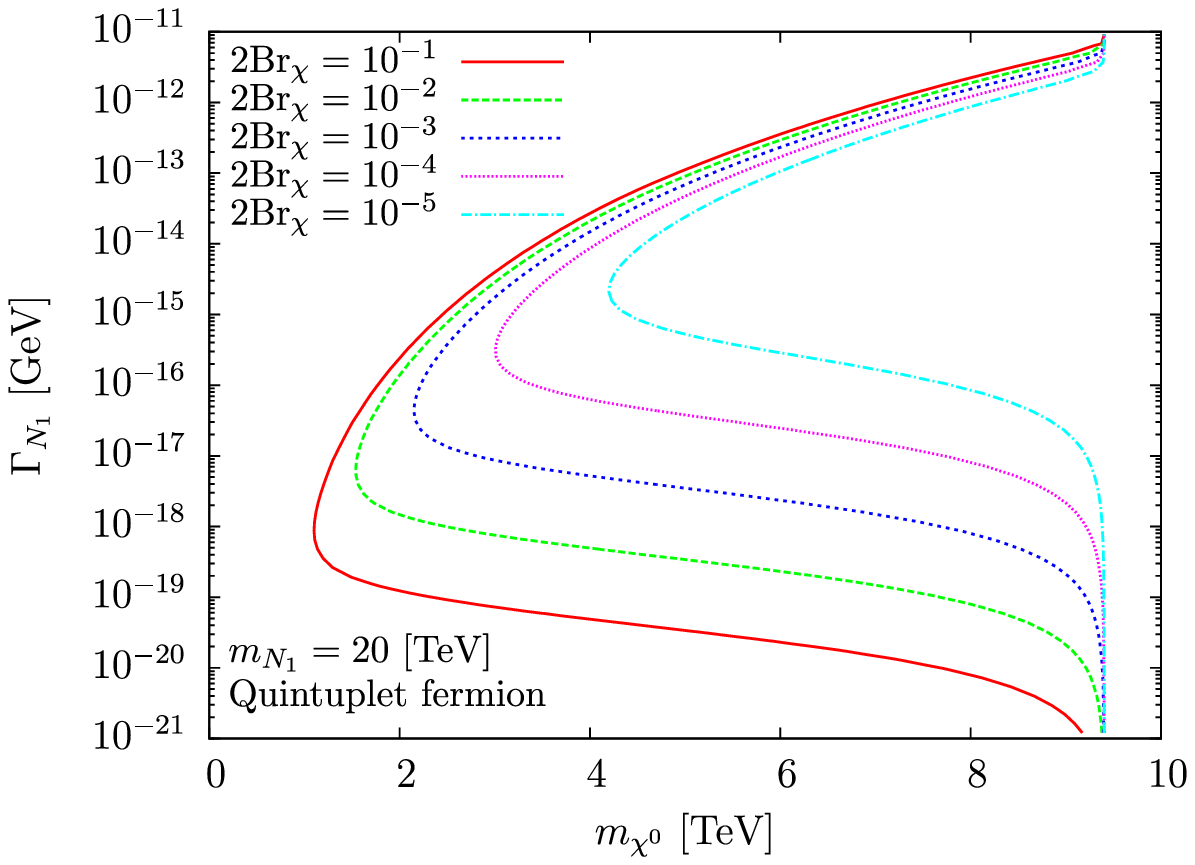}
\includegraphics[scale=0.57]{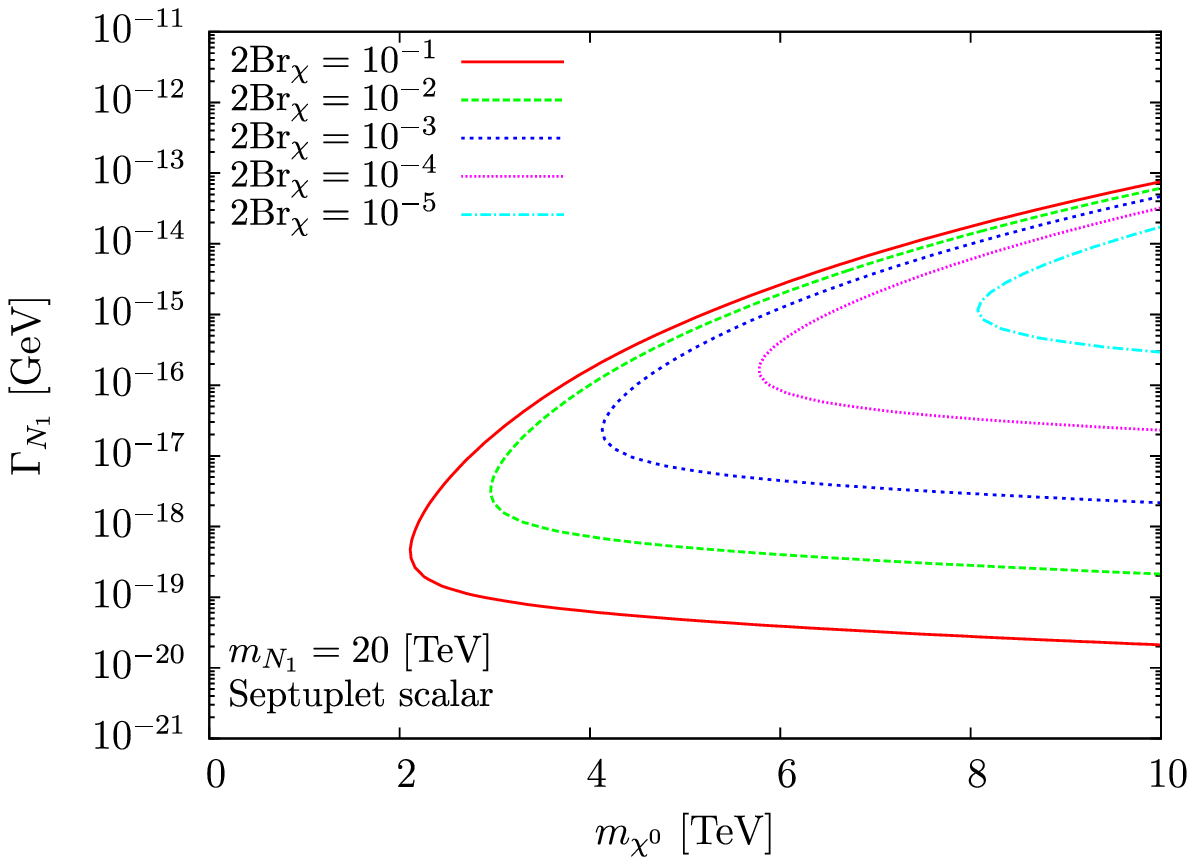}\\
\vspace{0.15cm}
\includegraphics[scale=0.57]{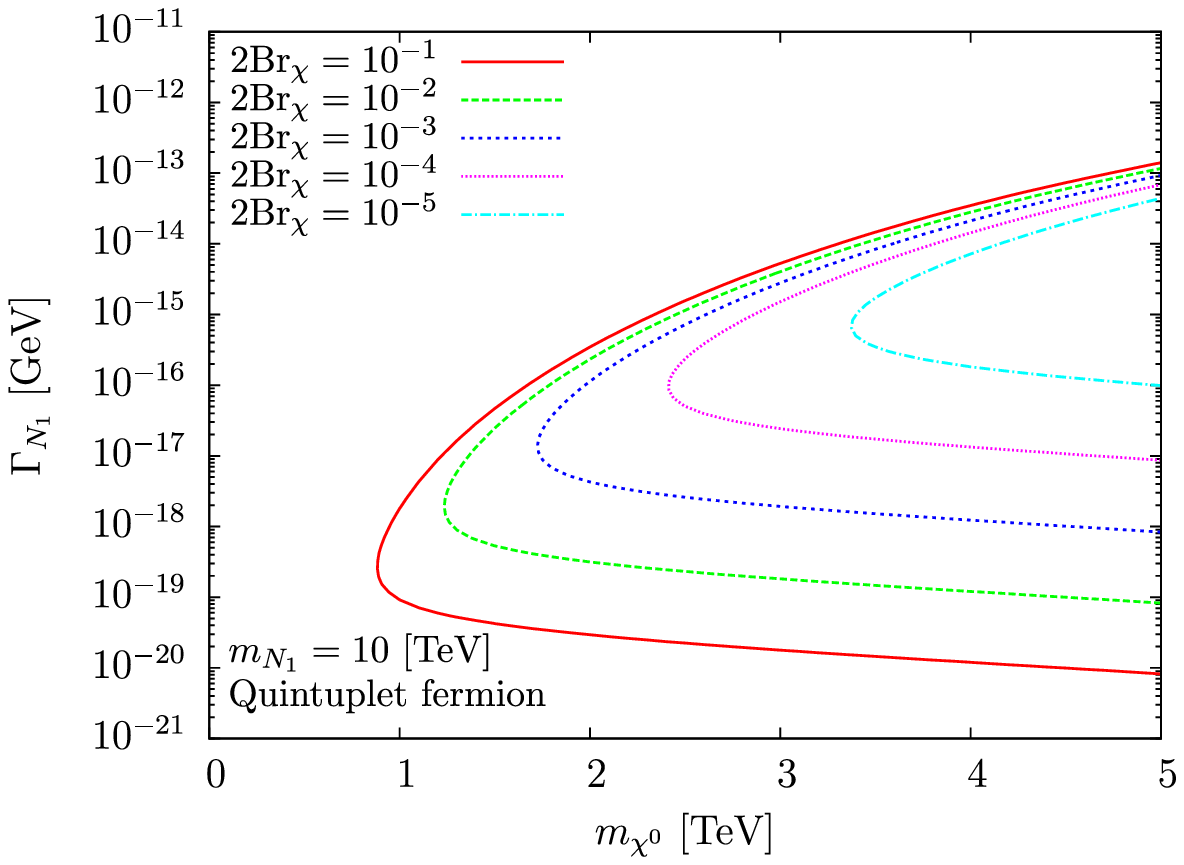}
\includegraphics[scale=0.57]{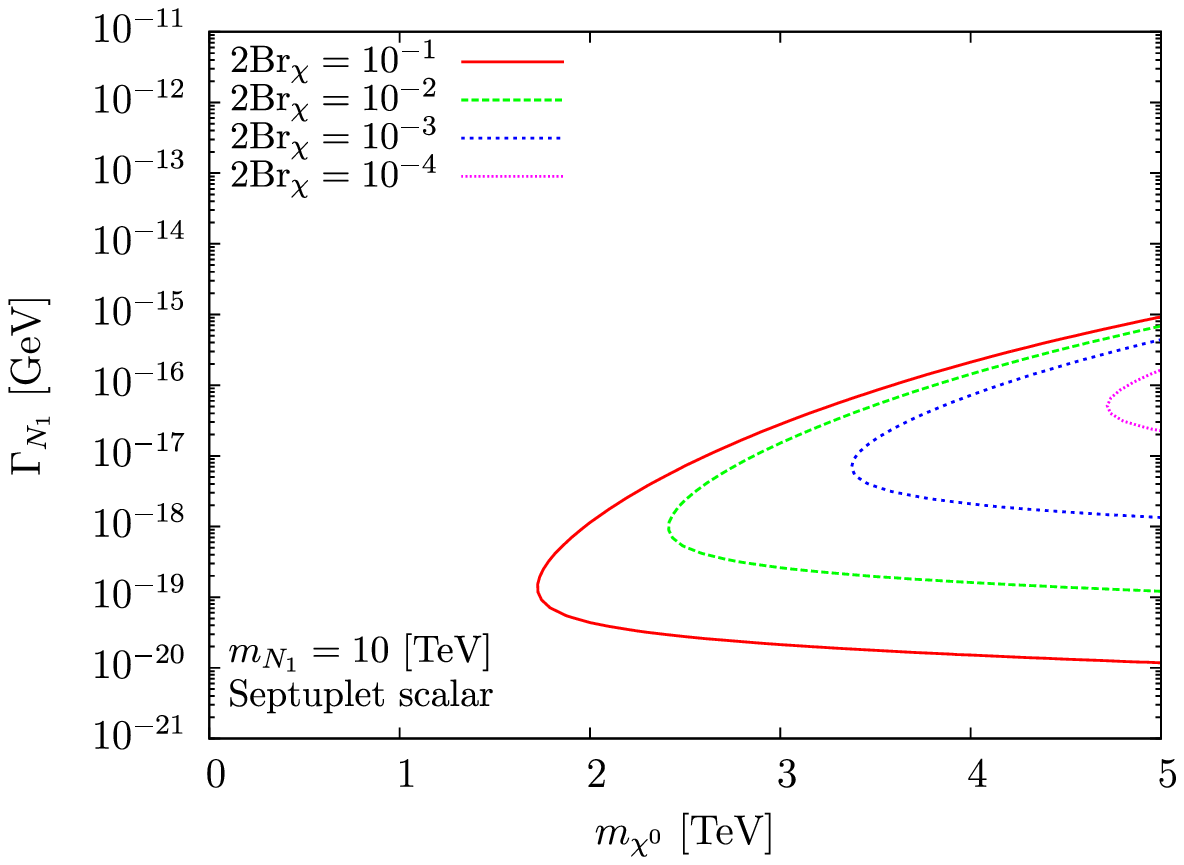}\\
\vspace{0.15cm}
\includegraphics[scale=0.57]{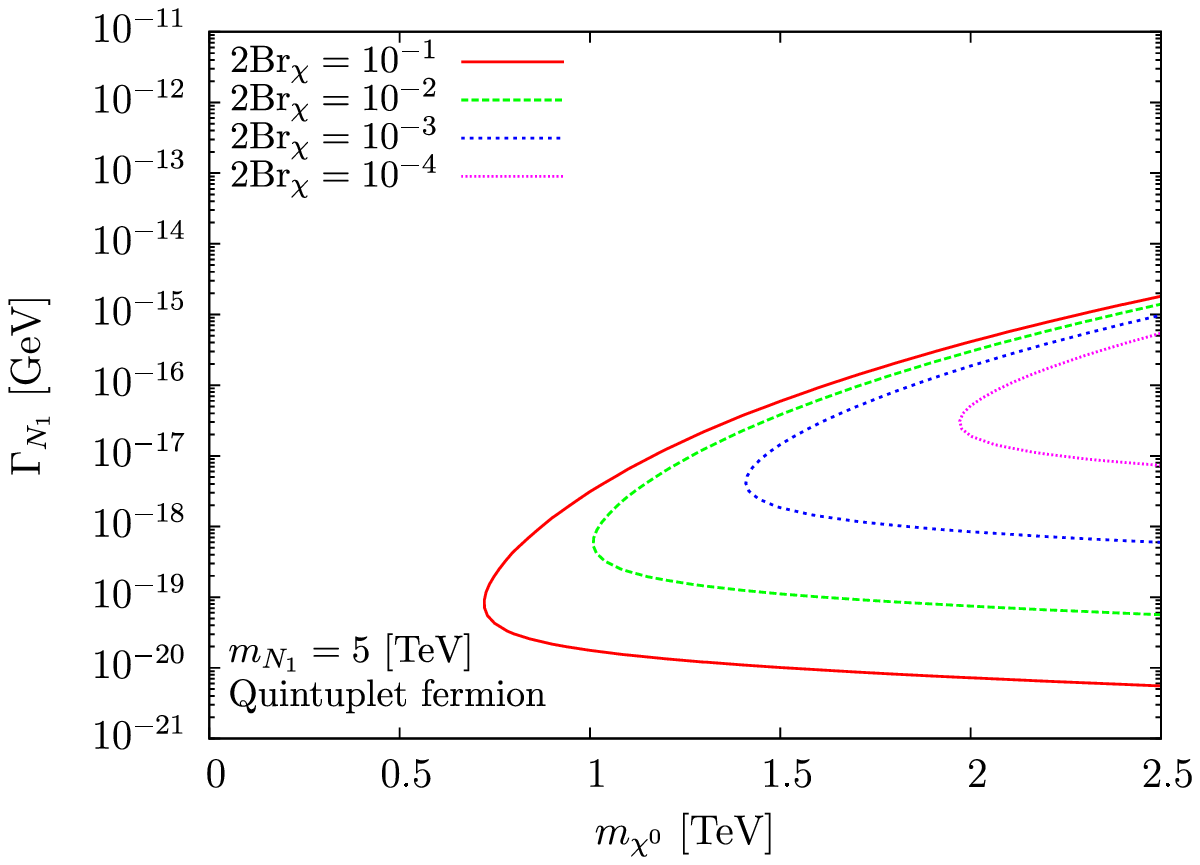}
\includegraphics[scale=0.57]{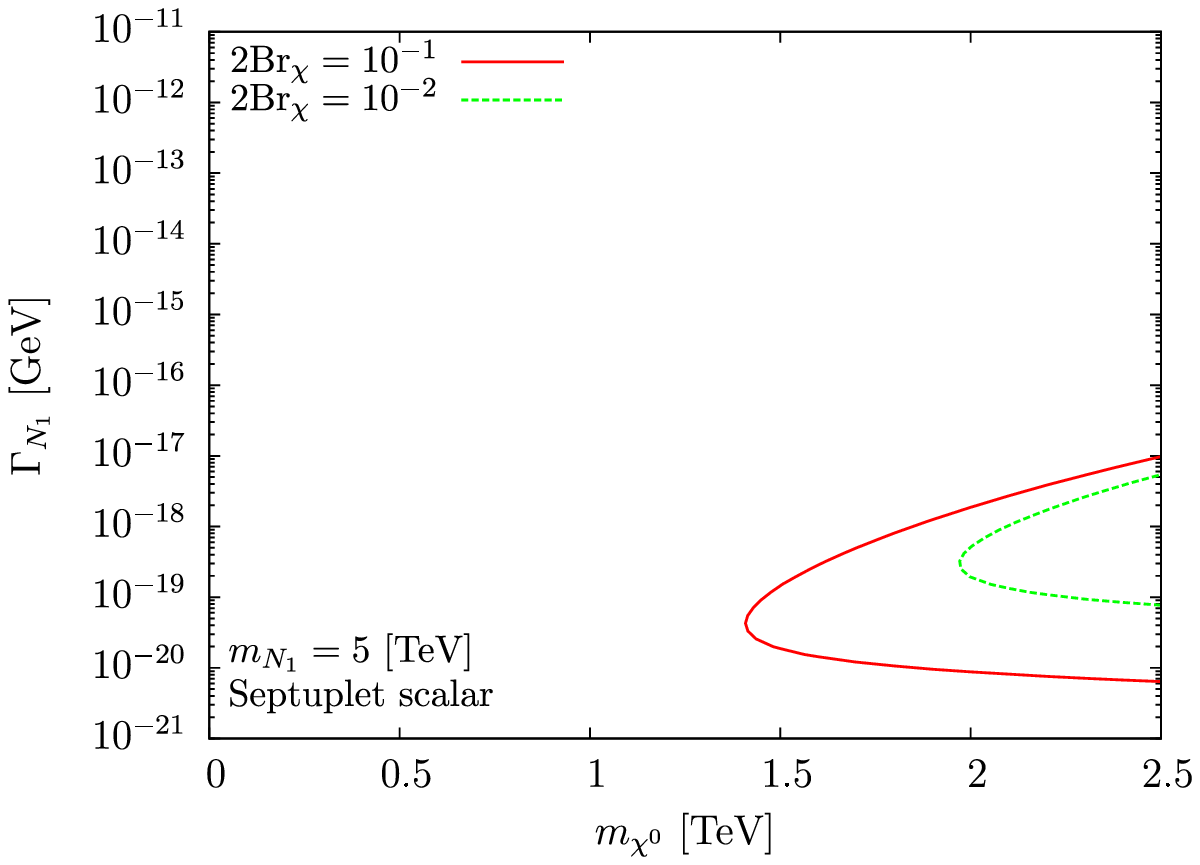}
\caption{Contours satisfying the observed DM relic density in the
  $m_{\chi^0}$-$\Gamma_{N_1}$ plane for several fixed
  $\mathrm{Br}_{\chi}$. The lightest right-handed neutrino mass is
  fixed to $m_{N_1}=20$ TeV for the upper panels,
 $m_{N_1}=10~\mathrm{TeV}$ for the middle ones and $m_{N_1}=5$ TeV
  for the bottom ones. The left panels are for the quintuplet fermion
  DM case while the right panels are for the septuplet scalar DM
  scenario. }
\label{fig:contour}
\end{center}
\end{figure}

The DM relic density is obtained by solving the Boltzmann equations.
For the numerical analysis, it proves convenient to replace the number
densities $n_{N_1}$ and $n_{\chi^0}$ by $Y_{N_1}=n_{N_1}/s$ and
$Y_{\chi^0}=n_{\chi^0}/s$, where $s$ is the entropy
density. Similarly, the time variable $t$ is rewritten in terms of the
temperature $T$.  Some examples of numerical solutions are shown in
Fig.~\ref{fig:boltzmann}, obtained for the quintuplet fermion DM
scenario, with $2 \, \mathrm{Br}_\chi = 0.001$, $m_{N_1} = 5$ TeV and
$m_{\chi^0} = 2$ TeV. For each fixed DM mass, two different values of
the decay width $\Gamma_{N_1}$ can satisfy the observed relic density.
For a small decay width $\Gamma_{N_1}$, the amount of produced $N_1$
is moderate and they slowly decay after DM freeze-out (left panel in
Fig.~\ref{fig:boltzmann}).  For a large decay width $\Gamma_{N_1}$, a
lot of $N_1$'s are created and they decay somewhat fast into DM
particles (right panel in Fig.~\ref{fig:boltzmann}).  In this case,
since the DM particles are still coupled with the thermal bath, the DM
production due to the decay of $N_1$ and the DM annihilation
compete. The correct relic density is finally obtained.

The parameter space satisfying the DM relic density measured by Planck
for several fixed values of $\mathrm{Br}_{\chi}$ is shown in
Fig.~\ref{fig:contour}.  One can see from the figure that for the
quintuplet fermion scenario, the decay width of the lightest heavy
neutrino $\Gamma_{N_1}$ should satisfy
\begin{eqnarray}
10^{-18}~\mathrm{GeV}\lesssim
\hspace{-0.2cm}&\Gamma_{N_1}&\hspace{-0.2cm}\lesssim10^{-12}~\mathrm{GeV}
\quad\mbox{for}\quad m_{N_1}=20~\mathrm{TeV},\nonumber\\
10^{-18}~\mathrm{GeV}\lesssim
\hspace{-0.2cm}&\Gamma_{N_1}&\hspace{-0.2cm}\lesssim10^{-13}~\mathrm{GeV}
\quad\mbox{for}\quad m_{N_1}=10~\mathrm{TeV},\nonumber\\
10^{-18}~\mathrm{GeV}\lesssim
\hspace{-0.2cm}&\Gamma_{N_1}&\hspace{-0.2cm}\lesssim10^{-15}~\mathrm{GeV}
\quad\mbox{for}\quad m_{N_1}=5~\mathrm{TeV},\nonumber
\end{eqnarray}
in order to get the measured DM relic density with $2 \,
\mathrm{Br}_{\chi}\lesssim0.001$.  Thus, one finds that the required
size of the lightest right-handed neutrino Yukawa coupling satisfies
$10^{-11}\lesssim y^\nu\lesssim10^{-8}$.  For the septuplet scalar DM
scenario, a larger branching ratio is required because the effective
annihilation cross section is larger than in the quintuplet fermion DM
case. In both cases we find that a much lower DM mass, compared to the
standard MDM scenario, is possible. This is the main result of our
paper.

A threshold at $m_{\chi^0}\approx9.4~\mathrm{TeV}$ is observed in the
upper left panel of Fig.~\ref{fig:contour}. When the DM mass is above
$9.4$ TeV, the effective annihilation cross section becomes too small,
leading to $\Omega_{\text{DM}} h^2 > 0.12$. Since the $N_1$ decay can
only increase the DM relic density, the $m_{\chi^0} >
9.4~\mathrm{TeV}$ region is excluded in our scenario. One should also
note that a lower bound on the decay width $\Gamma_{N_1}$ can be
derived from Big Bang Nucleosynthesis (BBN).  The conservative bound
for the $N_1$ lifetime $\tau_{N_1}\lesssim0.1~\mathrm{s}$, which
corresponds to $\Gamma_{N_1}\gtrsim10^{-23}~\mathrm{GeV}$, should be
taken into account not to affect the predictions of
BBN~\cite{Kawasaki:2004qu, Jedamzik:2006xz}.\footnote{If the $N_1$
  lifetime was longer than $10^3~\mathrm{s}$, the DM particles
  produced by the decay of $N_1$ would give an additional contribution
  to the number of effective neutrino species without affecting
  BBN. This is possible since the DM particles are relativistic due to
  their large kinetic energy~\cite{Hooper:2011aj}. However, this is
  not the case in our scenario, where the $N_1$ lifetime is shorter.}

Since the Yukawa coupling for the non-thermal production of
DM is required to be rather small, the lightest heavy neutrino $N_1$
does not play any role in the generation of active neutrino masses. As
a result of this, the active neutrino masses and mixings are explained
by the other heavy neutrinos $N_2$ and $N_3$, with the Yukawa
couplings of the order of $y^\nu\sim10^{-6}$.

We briefly comment on DM particles lighter than the $W$ boson mass.
For DM masses below $m_W$, the DM relic density can be non-thermally
produced with $2\mathrm{Br}_\chi\gtrsim0.01$ if the mass of the
lightest right-handed neutrino is $m_{N_1}\lesssim 300~\mathrm{GeV}$.
However, there are two reasons to disregard this scenario. First, the
required branching ratio into DM is too large and calls for an
extension of the model, as explained below. And second, and more
importantly, this DM mass range is completely excluded by the
constraints from collider experiments as discussed in
Sec.~\ref{sec:constraints-collider}.

\subsubsection*{Getting the required $\mathrm{Br}_{\chi}$ value}

In the minimal models discussed in this paper one expects a small
$\mathrm{Br}_{\chi}$ value for $m_{\chi^0}\gg m_W,m_Z$.  The main
reason is intuitive: three body decays are suppressed compared to two
body ones since the mediators of the three body decays, the $W^\pm$ and
$Z$ gauge bosons, are much lighter than $N_1$. 
In the septuplet scalar case, the Higgs boson can also mediate these
decays via the scalar coupling $\lambda_{\phi\chi}$. However, the
branching ratio $\mathrm{Br}_\chi$ cannot be made as large as required to
account for the DM relic density even if the maximum
$\lambda_{\phi\chi}$ value allowed by direct detection experiments is
considered.

This problem can be easily solved by adding a new heavy mediator for
the three body decays. The simplest one would be a real scalar
$\sigma$, with couplings to a pair of $\chi$ fields ($y_\sigma m_\chi \sigma
\chi \chi$ or $y_\sigma\sigma \overline{\chi^c} \chi$ depending on the
variant of the MDM scenario considered).  The real scalar $\sigma$
would mix with the SM Higgs boson $h$ and, as a consequence of this,
the lightest right-handed neutrino $N_1$ would decay into
$N_1\to\chi\overline{\chi}\nu_\alpha$.  One should note that although
the term $\sigma \overline{N_i^c}N_j$ is also possible, this coupling
should be small enough so that $N_1$ is not in thermal equilibrium and
can be produced by the freeze-in mechanism.  We find that the required
value of the $h-\sigma$ mixing angle is $\sin\alpha\sim0.1$ and the
mass of the new mediator ($H$, the heaviest mass eigenstate resulting
from $h$ and $\sigma$) is $m_H\gtrsim m_\chi$ for
$y_\sigma=\mathcal{O}(0.1)$.  The coupling $y_\sigma$ is also
constrained by DM direct detection experiments, and we have checked
that the value $y_\sigma=\mathcal{O}(0.1)$ is consistent with the
current LUX bound for TeV scale DM.  We also have checked that the
coupling $y_\sigma=\mathcal{O}(0.1)$ does not affect to the DM relic
density and indirect detection, which will be discussed in the
following section, since the electroweak interaction of the MDM
multiplet is dominant.  The required mixing angle is perfectly
consistent with the current measurements of the Higgs properties by
the LHC \cite{Robens:2015gla, Bonilla:2015uwa}.  Finally, such a real
scalar could be related to the breaking of an extra symmetry existing
at higher energy scales such as, for example, $U(1)_{L}$ or
$U(1)_{B-L}$, where $L$ and $B$ are lepton and baryon numbers,
respectively.

\section{Constraints}
\label{sec:constraints}

In this section we review the most relevant constraints in our
scenario. These come from collider searches as well as from dark
matter indirect and direct detection experiments.

\subsection{Collider constraints in low dark matter mass scenarios}
\label{sec:constraints-collider}

\begin{figure}[t]
\begin{center}
\includegraphics[scale=0.65]{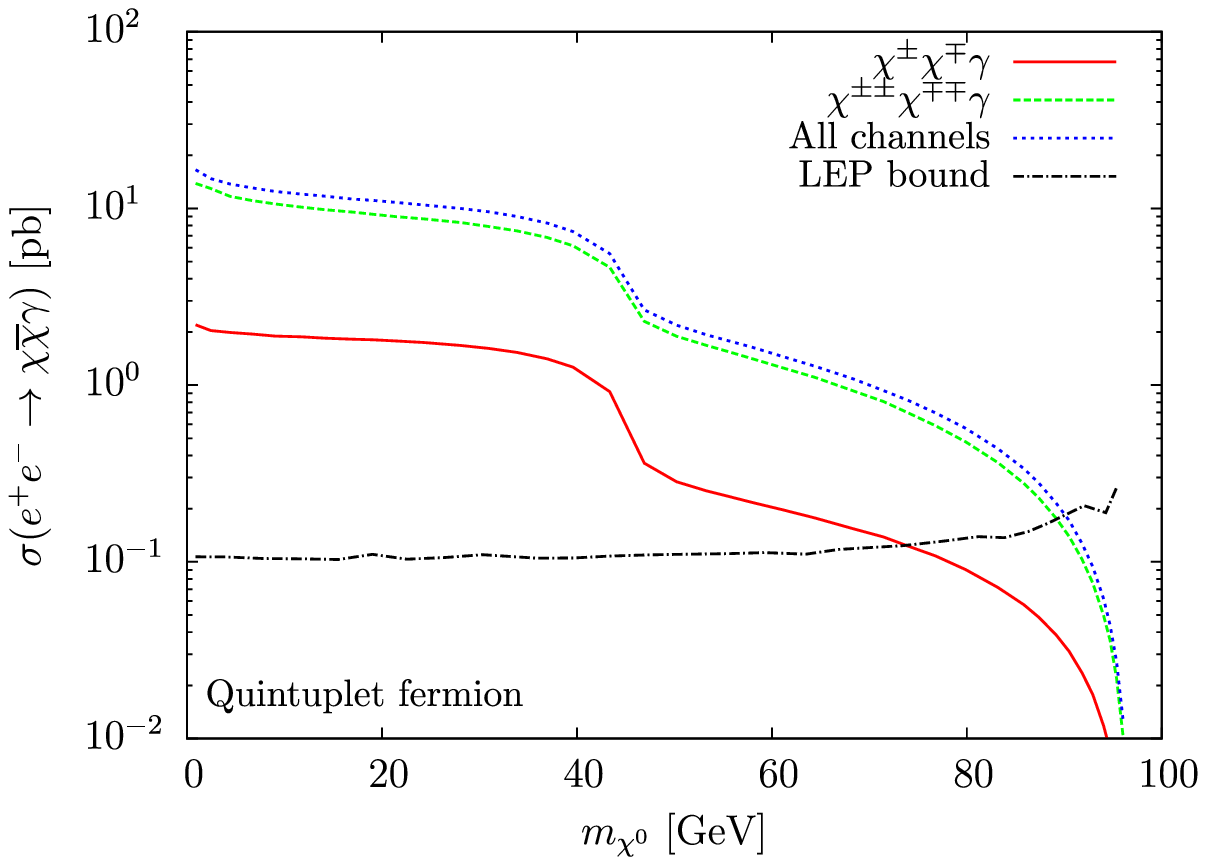}
\includegraphics[scale=0.65]{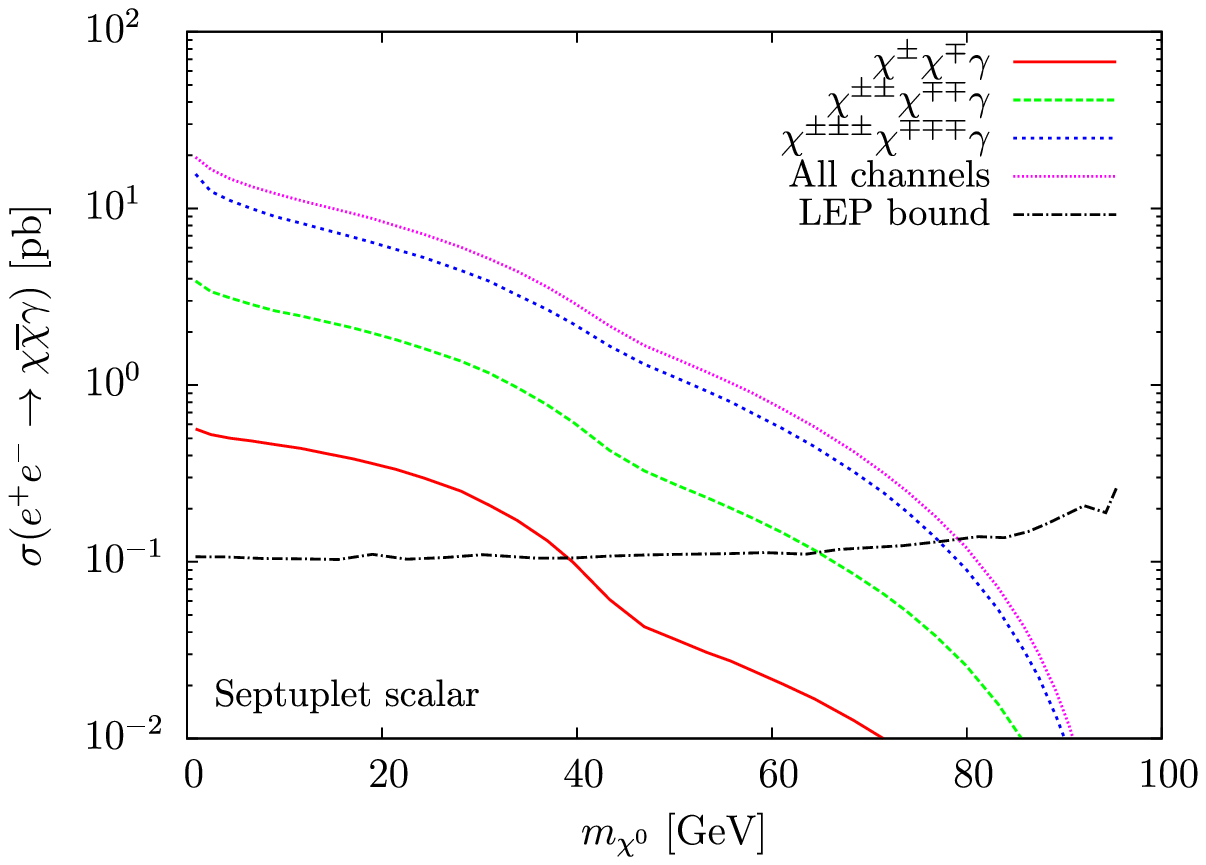}
\caption{LEP limits on quintuplet fermion (left) and septuplet scalar
  (right) scenarios. In the derivation of these limits we have taken
  photon polar angles in the $46^\circ<\theta<135^\circ$ range and
  photon energies $E_\gamma>6~\mathrm{GeV}$.}
\label{fig:lep}
\end{center}
\end{figure}

Let us briefly comment on collider constraints in our scenario. These
are of course only relevant for low dark matter masses and, as we will
see, they strongly restrict this possibility.

First, we consider LEP bounds. Collider constraints derived from
mono-photon searches at LEP turn out to be very strong for low dark
matter masses.  Multi-charged particles plus a photon are produced by
the process
\begin{equation*}
e^+e^- \to
 Z^*/\gamma^*\to\chi^{\pm}\chi^{\mp}\gamma,
~\chi^{\pm\pm}\chi^{\mp\mp}\gamma,
~\Bigl(\chi^{\pm\pm\pm}\chi^{\mp\mp\mp}\gamma\Bigr)
 \, .
\end{equation*}
Then, the multi-charged particles decay into
\begin{eqnarray*}
&&\chi^{\pm}\to
\chi^0W^{\pm*} \to\chi^0\pi^\pm \, ,\\
&&\chi^{\pm\pm}\to
\chi^0W^{\pm*} W^{\pm*}\to\chi^0\pi^\pm\pi^\pm \, ,\\
&&\chi^{\pm\pm\pm}\to
\chi^0W^{\pm*} W^{\pm*} W^{\pm*}\to\chi^0\pi^\pm\pi^\pm\pi^\pm \, ,
\end{eqnarray*}
producing soft pions due to the mass degeneracy among the components
of the MDM multiplet $\chi$. Therefore, if these pions are not seen
due to their low energies, the resulting signal at LEP is mono-photon
plus missing energy $e^+e^-\to\gamma E\!\!\!\!/\:$. Using DELPHI
data~\cite{Abdallah:2003np, Abdallah:2008aa}, the authors of Ref.~\cite{Fox:2011fx}
used this idea to set limits on the suppression scale $\Lambda$ of
four-fermion contact interactions of the type $\mathcal{O}_4 \sim
\frac{1}{\Lambda^2} \bar \chi \chi \, \bar e \, e$.

Here, we translate the lower limits of the suppression scale $\Lambda$
into upper limits on the total production cross section for
$e^+e^-\to\chi\overline{\chi}\gamma$, where $\chi$ denotes $\chi^\pm$,
$\chi^{\pm\pm}$ ($\chi^{\pm\pm\pm}$).  Although the energy 
distribution of the mono-photon in our case is not exactly the same as
that obtained with the contact interactions, this estimate should
provide a rough limit in our scenario.
For our analysis we consider photons with polar angles $\theta$ in the
range $45^\circ<\theta<135^\circ$ and energies
$E_\gamma>6~\mathrm{GeV}$. These kinematical cuts are required to be
consistent with the detection capabilities of the High Density
Projection Chamber of DELPHI. Our results are shown in
Fig.~\ref{fig:lep}, both for the quintuplet fermion and septuplet
scalar. As one can see from the figures, a slightly stronger
constraint is obtained for the quintuplet fermion.  DM masses
$m_{\chi^0}\lesssim 90~\mathrm{GeV}$ for the quintuplet fermion and
$m_{\chi^0}\lesssim79~\mathrm{GeV}$ for the septuplet scalar are
excluded.  We point out that very similar bounds have been 
obtained in the recent Ref.~\cite{DiLuzio:2015oha}.

Regarding LHC bounds, these have been recently analyzed in
\cite{Ostdiek:2015aga}. For the quintuplet fermion case, the lower bound
$m_{\chi^0}\gtrsim267~\mathrm{GeV}$ has been obtained with the LHC
running at $\sqrt{s}=8~\mathrm{TeV}$ and $20.3~\mathrm{fb}^{-1}$ of
integrated luminosity in ATLAS and $19.5~\mathrm{fb}^{-1}$ in
CMS~\cite{Ostdiek:2015aga}.
This bound will be improved up to
$m_{\chi^0}\gtrsim668~\mathrm{GeV}$ if the LHC does not find a
signal with $\sqrt{s}=14~\mathrm{TeV}$ and $3~\mathrm{ab}^{-1}$.

\subsection{Gamma-ray constraints}

The DM annihilation channels $\chi^0\chi^0\to W^+W^-$ and
$\chi^0\chi^0\to\gamma\gamma$ induce indirect detection signals of DM.
At present time, the cross sections for these two annihilation
processes are drastically affected by the non-perturbative Sommerfeld
effect due to the low kinetic energy of the DM particles. This
typically leads to relevant constraints from observations of
gamma-rays coming from dwarf spheroidal satellite galaxies or the
galactic center~\cite{Abramowski:2013ax,Ackermann:2015zua}.  Note that
even for the $W^+W^-$ annihilation channel, high energy gamma-rays are
generated in the decay of the $W$ boson.

Let us elaborate on the so-called Sommerfeld effect.  As we have
already discussed, the components of the MDM multiplets are naturally
degenerate, and the mass scale of the DM particles is much higher than
that of the gauge boson masses. In this case, the usual perturbative
calculation for annihilation cross sections is not valid because
long-range Coulomb-like forces which imply the Yukawa forces with 
small mediator masses among the MDM components distort the
plane wave function of the incoming DM two-body state.  Hence, the
annihilation cross sections must be calculated non-perturbatively by
taking into account the Sommerfeld
correction~\cite{Sommerfeld:1931}. This was pointed out for the first
time for a wino-like neutralino DM scenario in supersymmetric
models~\cite{Hisano:2002fk,Hisano:2003ec,Hisano:2004ds,
  Hisano:2005ec,Hisano:2006nn}.
A similar calculation was performed for MDM models in
Ref.~\cite{Cirelli:2007xd}.  In our models, the calculation is basically 
the same, except for the mass range we focus on. Therefore, we must
proceed to the re-evaluation of the Sommerfeld enhanced cross
sections, in order to be able to compare to the current bounds from
indirect detection experiments.

Our numerical analysis follows Ref.~\cite{Baumgart:2014saa} and
includes only the $s$-wave component of the cross section. In order to
obtain the correction factor for the amplitude induced by the
Sommerfeld effect, we must solve a coupled Schr\"odinger equation in
the presence of a potential generated by long-range forces,
\begin{equation}
-\frac{1}{m_{\chi^0}}\frac{d^2\psi_i}{dr^2}+V_{ij}\psi_j=\frac{m_{\chi^0} v^2}{4}\psi_i,
\end{equation}
where $\psi_i$ is the wave function of the two-body DM state and
$V_{ij}$ is the potential matrix. The indices $i,j$ run as $i,j=1-3$
for the quintuplet fermion and as $i,j=1-4$ for the septuplet scalar.
The wave function $\psi$ and the potential matrix $V$ are explicitly
given by
\begin{equation}
\psi=\left(
\begin{array}{c}
\langle r|\chi^{++}\chi^{--}\rangle\\
\langle r|\chi^{+}\chi^{-}\rangle\\
\langle r|\chi^{0}\chi^{0}\rangle
\end{array}
\right),\qquad
V=\left(
\begin{array}{ccc}
8\Delta-4A & -2B & 0\\
-2B & 2\Delta-A & -3\sqrt{2}B\\
0 & -3\sqrt{2}B & 0
\end{array}
\right),
\end{equation}
for the quintuplet fermion and 
\begin{equation}
{\bf \psi}=\left(
\begin{array}{c}
\langle r|\chi^{+++}\chi^{---}\rangle\\
\langle r|\chi^{++}\chi^{--}\rangle\\
\langle r|\chi^{+}\chi^{-}\rangle\\
\langle r|\chi^{0}\chi^{0}\rangle
\end{array}
\right)\qquad
V=\left(
\begin{array}{cccc}
18\Delta-9A & -3B & 0 & 0\\
-3B & 8\Delta-4A & -5B & 0\\
0 & -5B & 2\Delta-A & -6\sqrt{2}B\\
0 & 0 & -6\sqrt{2} & 0
\end{array}
\right),
\end{equation}
for the septuplet scalar. Here
$A=\alpha_\mathrm{em}/r+\alpha_W\cos^2\theta_We^{-m_Zr}/r$,
$B=\alpha_We^{-m_Wr}/r$, $\Delta=166~\mathrm{MeV}$ and $r$ is the
distance between the two DM particles.  The non-trivial factor
$\sqrt{2}$ appears in some matrix elements of the potential due to the
different normalization between the neutral and charged states.

We are interested in the annihilation of the neutral state, which is
governed by physics at short distances and thus described by the wave
function at the origin, $\psi(0)$. Therefore, as discussed in
Refs.~\cite{Cirelli:2007xd, ArkaniHamed:2008qn, Baumgart:2014saa}, in
order to evaluate the Sommerfeld correction factor, we must
solve the Schr\"odinger equation with linear independent boundary
conditions for irregular solutions at the origin.  For the quintuplet
fermion, these boundary conditions at the origin and at $r\to\infty$
are explicitly given by
\begin{equation}
\psi(0)=\left(
\begin{array}{c}
1\\
0\\
0
\end{array}
\right),\quad
\left(
\begin{array}{c}
0\\
1\\
0
\end{array}
\right),\quad
\left(
\begin{array}{c}
0\\
0\\
1
\end{array}
\right),\quad
\frac{d\psi(\infty)}{dr}=\left(
\begin{array}{ccc}
ik_1 & 0 & 0\\
0 & ik_2 & 0\\
0 & 0 & ik_3
\end{array}
\right)\psi(\infty) \, ,
\end{equation}
where $k_i\equiv\sqrt{m_{\chi^0}^2v^2-m_{\chi^0} V_{ii}(\infty)}/2$.  After
solving the Schr\"odinger equation for each boundary condition at the
origin, the Sommerfeld factor matrix $A_{ij}$ is given by
\begin{equation}
A_{ij}=\left.\psi_{i}^{(j)}(r)e^{-ik_ir}\right|_{r\to\infty} \, ,
\end{equation}
where the superscript $j$ implies $j$-th boundary condition.  The
above discussion is straightforwardly extended to the septuplet
scalar.

Finally, the absorptive parts describing the DM annihilations for the
$W^+W^-$ and $\gamma\gamma$ channels are given by
\begin{equation}
\Gamma_{WW}=\frac{\pi\alpha_W^2}{m_{\chi^0}^2}\left(
\begin{array}{ccc}
2 & 5 & 3\sqrt{2}\\
5 & 25/2 & 15/\sqrt{2}\\
3\sqrt{2} & 15/\sqrt{2} & 9
\end{array}
\right),\quad
\Gamma_{\gamma\gamma}=\frac{\pi\alpha_{\mathrm{em}}^2}{m_{\chi^0}^2}\left(
\begin{array}{ccc}
16 & 4 & 0\\
4 & 1 & 0\\
0 & 0 & 0
\end{array}
\right),
\end{equation}
for the quintuplet fermion and 
\begin{equation}
\Gamma_{WW}=\frac{\pi\alpha_W^2}{m_{\chi^0}^2}\left(
\begin{array}{cccc}
9 & 24 & 33 & 18\sqrt{2}\\
24 & 64 & 88 & 48\sqrt{2}\\
33 & 88 & 121 & 66\sqrt{2}\\
18\sqrt{2} & 48\sqrt{2} & 66\sqrt{2} & 72
\end{array}
\right),\quad
\Gamma_{\gamma\gamma}=\frac{2\pi\alpha_{\mathrm{em}}^2}{m_{\chi^0}^2}\left(
\begin{array}{cccc}
81 & 36 & 9 & 0\\
36 & 16 & 4 & 0\\
9 & 4 & 1 & 0\\
0 & 0 & 0 & 0
\end{array}
\right),
\end{equation}
for the septuplet scalar.  The DM annihilation cross sections can be
computed from the absorptive parts and the Sommerfeld factor $A$ as
\begin{equation}
\sigma{v}_{WW}=2\left(A\Gamma_{WW}A^\dag\right)_{33},\qquad
\sigma{v}_{\gamma\gamma}=2\left(A\Gamma_{\gamma\gamma}A^\dag\right)_{33},
\end{equation}
for the quintuplet fermion and 
\begin{equation}
\sigma{v}_{WW}=2\left(A\Gamma_{WW}A^\dag\right)_{44},\qquad
\sigma{v}_{\gamma\gamma}=2\left(A\Gamma_{\gamma\gamma}A^\dag\right)_{44},
\end{equation}
for the septuplet scalar. 

\begin{figure}[t]
\begin{center}
\includegraphics[scale=0.65]{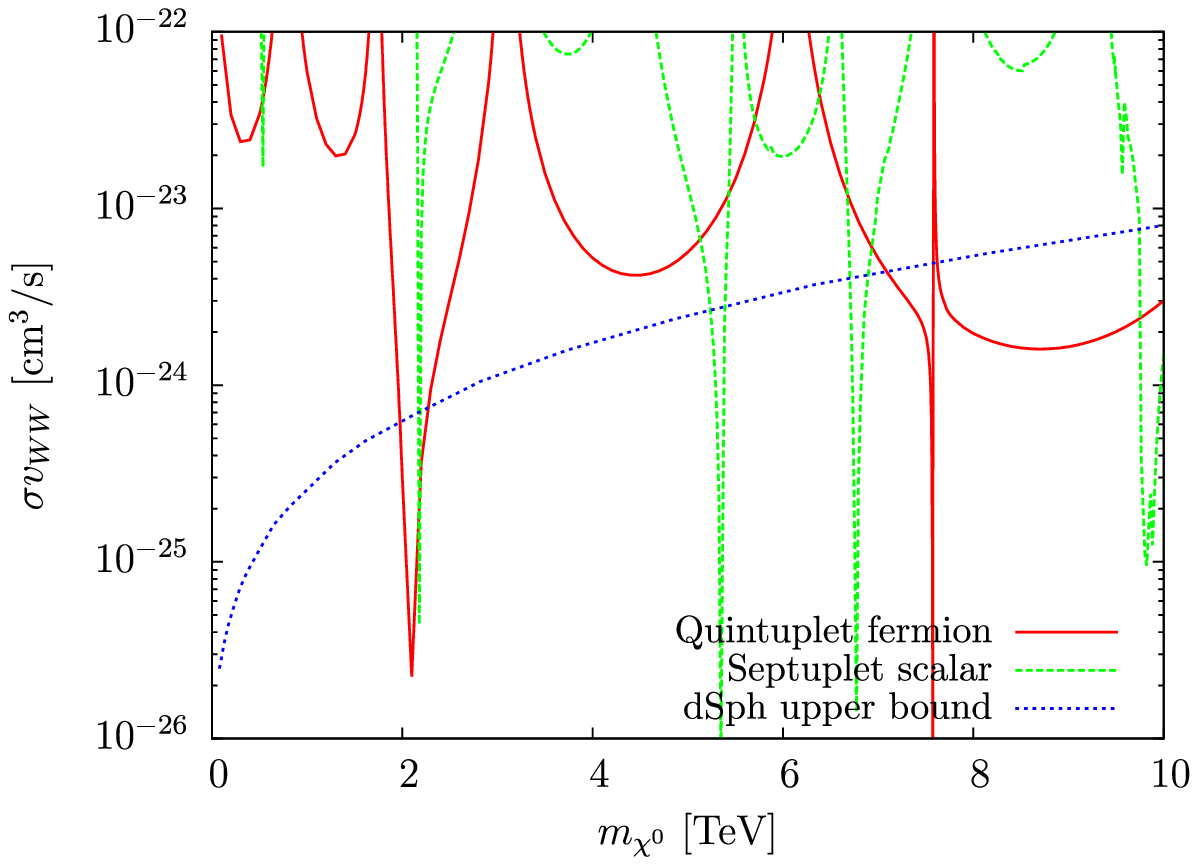}
\includegraphics[scale=0.65]{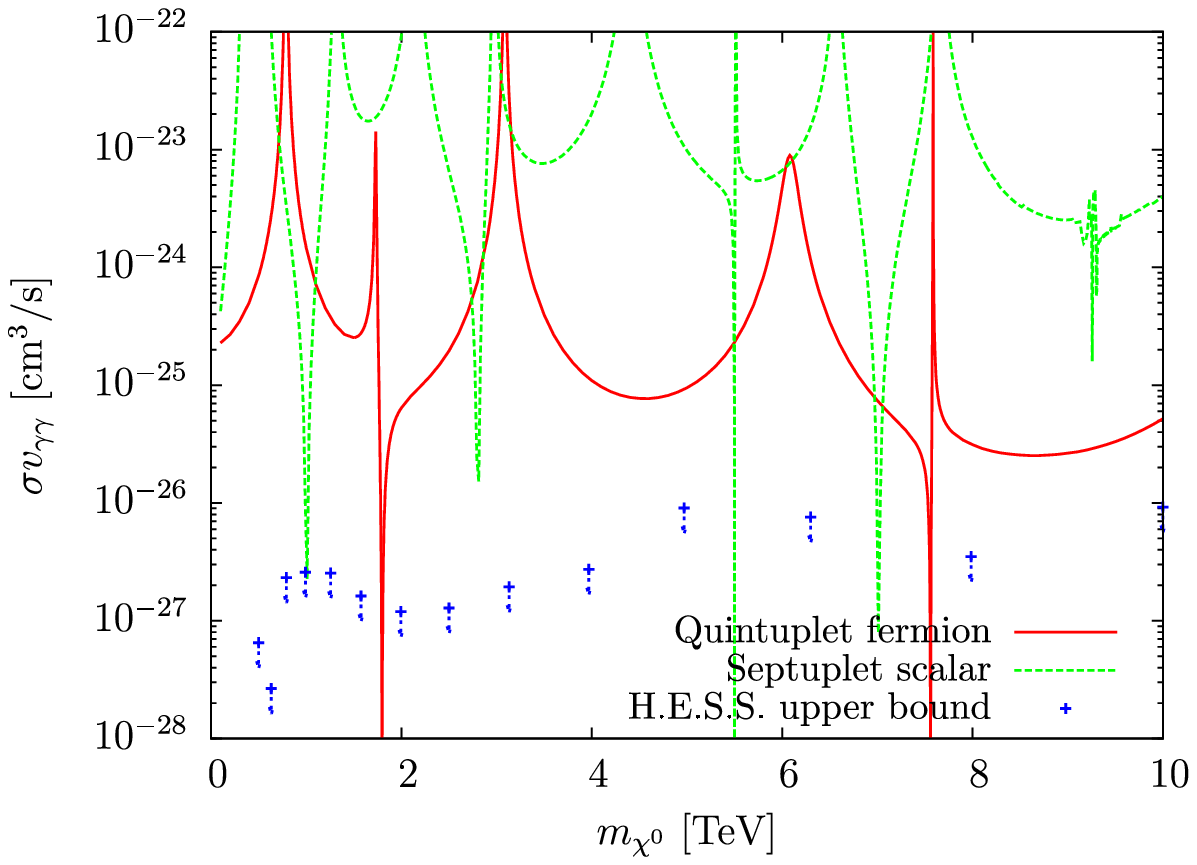}
\caption{Cross sections for $\chi^0\chi^0\to W^+W^-$ and
  $\chi^0\chi^0\to\gamma\gamma$. The DM relative velocity is fixed to
  $v=10^{-3}$.}
\label{fig:sommerfeld}
\end{center}
\end{figure}

Our numerical results for the quintuplet fermion and septuplet scalar
into $W^+W^-$ and $\gamma\gamma$ are shown in
Fig.~\ref{fig:sommerfeld}. The DM relative velocity is fixed to
$v=10^{-3}$. Since the actual DM velocity has a distribution, the
cross section should be averaged over the DM velocity in a more
sophisticated analysis. For the $W^+W^-$ channel, the upper bound on
gamma-ray observations coming from dwarf spheroidal galaxies (assuming
NFW profile) is also shown~\cite{Ackermann:2015zua}.  For the
$\gamma\gamma$ channel, the blue points show the H.E.S.S. limit for
monochromatic gamma-ray lines coming from the galactic centre,
assuming in this case a Einasto profile~\cite{Abramowski:2013ax}. As
one can see from the figures, the cross sections largely exceed the
bounds except in several dips.  These dips can be interpreted in
analogy with the Ramsauer-Townsend effect, the scattering of low
energy electrons by atoms of a noble gas~\cite{Chun:2012yt,
Chun:2015mka}. 
The Ramsauer-Townsend effect is caused by the electromagnetic interaction,
and the positions of the dips strongly depend on the mass difference
$\Delta$~\cite{Chun:2012yt, Chun:2015mka}.  Due to this effect, the incoming two DM
particles pass through unaffected by the potential when the DM pair
has a fixed energy, thus drastically decreasing the annihilation cross
section.  When the mass difference $\Delta$ is as large as
$\mathcal{O}(10)$ GeV, the dips tends to disappear since the
electromagnetic transition between the DM and the charged states
becomes unefficient.  The resonant behaviour in
Fig.~\ref{fig:sommerfeld} is caused by the bound state of the two DM
particles with zero binding energy~\cite{Hisano:2003ec, Hisano:2004ds,
  Chun:2012yt, Chun:2015mka}.

For the quintuplet fermion, the dips appear at DM masses around
$2~\mathrm{TeV}$ and $7.5~\mathrm{TeV}$ for the $W^+W^-$ channel.  The
positions of the dips are very close to those in the $\gamma\gamma$
channel.  This is the only possibility to evade the gamma-ray
constraints simultaneously for $W^+W^-$ and $\gamma\gamma$ final
states.  For the septuplet scalar, the same thing occurs at around
$5.5~\mathrm{TeV}$ and $7~\mathrm{TeV}$.

Note that the gamma-ray constraints depend on DM profiles.  If a cored
DM profile such as Burkert or Isothermal is assumed, these constraints
are relaxed. Concerning this uncertainty, two orders of magnitude looser
bounds are expected at most.  Moreover since the positions of the dips
due to the Ramsauer-Townsend strongly depend on the mass difference
$\Delta$, if a larger mass difference, like
$\mathcal{O}(1)~\mathrm{GeV}$, were possible by an extension of the
model, a broader DM mass range would be able to satisfy the gamma-ray
constraints.

We point out that for the septuplet scalar DM, there are
additional contributions to the annihilations into $W^+W^-$ and
$\gamma\gamma$ due to the scalar couplings $\lambda_{\phi\chi}$ and
$\lambda_\chi$.  These contributions would lead to an increase. However,
they are found to be subdominant compared to the gauge boson loops (with
the Sommerfeld correction calculated above).

One should note that the above discussion of the dips
due to the Ramsauer-Townsend effect is controversial. More detailed
analyses have recently been done in Refs.~\cite{Cirelli:2015bda,
Garcia-Cely:2015dda}. According to these references, in addition
to the $W^+W^-$ and $\gamma\gamma$ channels, gamma-ray production in
the $ZZ$, $Z\gamma$ and $W^+W^-\gamma$ channels should also be taken
into account.  If these channels are included, the dips disappear
and thermally produced MDM with a mass
$m_{\chi^0}\approx9.4~\mathrm{TeV}$ for the quintuplet fermion or
$m_{\chi^0}\approx25~\mathrm{TeV}$ for the septuplet scalar is
excluded in the case of cusp DM profiles such as NFW and Einasto,
but still allowed for cored profiles like
Isothermal.\footnote{Another way to evade the gamma-ray constraint 
would be to assume a sub-dominant MDM scenario.} For lower DM
masses, as in our case, the gamma-ray constraint becomes stronger
than in the thermal MDM scenario. There is, however, a valid region
in the parameter space (with a specific DM mass) when cored DM
profiles are considered.

\subsection{Direct detection}

In the MDM scenario, elastic scattering with quarks is induced at the
one-loop level via $SU(2)_L$ gauge interactions. Our setup with
non-thermal DM production allows for lighter DM particles than in the
in the standard MDM scenario with thermally produced DM.  With
non-thermal production we find valid DM masses below
$7.5~\mathrm{TeV}$ in both cases, whereas in the standard thermal
scenario one has $9.4~\mathrm{TeV}$ for the quintuplet fermion and
$25~\mathrm{TeV}$ for the septuplet scalar. For this reason, we expect
that the constraints derived from direct detection experiments will be
stronger than those obtained in the usual thermal scenario.

The one-loop spin independent elastic cross section with a proton was
found in Ref.~\cite{Cirelli:2005uq} to be about
$\sigma_p\sim10^{-44}~\mathrm{cm^2}$. This value for the cross section
may seem too large compared to the current experimental bound obtained
by LUX for a DM mass $m_{\chi^0}\lesssim7.5~\mathrm{TeV}$.  However,
according to recent calculations including two-loop diagrams including
DM and gluons, the Higgs mass measured at the LHC and recent lattice
simulations for the strangeness content of the nucleon, it turns out
that the elastic cross section gets reduced due to partial
cancellations, leading to
$\sigma_{p}\sim10^{-46}~\mathrm{cm^2}$~\cite{Farina:2013mla,
  Hisano:2015rsa}. This is below the current LUX bound and testable by
the future direct detection experiments such as XENON1T.  Given that
the DM particles in our scenario are lighter than those present in
scenarios with thermally produced MDM, the coming direct detection
experiments will also test our setup.

\section{Summary and Conclusions}
\label{sec:sum}

In this paper we have discussed an extension of the SM with three
right-handed neutrinos and a large $SU(2)_L$ multiplet. The $SU(2)_L$
multiplet is either a quintuplet fermion or a septuplet scalar.
Despite imposing no additional symmetry, the lightest neutral
component of the multiplet can constitute the DM content of the
universe because of an accidental symmetry, as in the conventional MDM
scenario.  Furthermore, neutrino masses are induced by the
canonical Type-I seesaw mechanism.

However, unlike the conventional MDM scenario with thermally produced
DM, in our setup the DM particles are non-thermally produced by the
decay of the heavy neutrinos. This allows to lower significantly the
DM mass and still be compatible with the observed DM relic
density. Instead of DM masses as large as $9.4$ TeV or $25$ TeV, the
DM mass in our non-thermal scenario can be as light as a few TeV.

Finally, we have considered several experimental constraints in our
scenario. First, we have discussed the possibility of a DM mass below
the $W$ boson mass, which is excluded due to strong constraints coming
from mono-photon plus missing energy searches at LEP. In particular,
we found that the mass ranges $m_{\chi^0} \lesssim 90$ GeV and
$m_{\chi^0} \lesssim 79$ GeV are excluded for the quintuplet fermion
and septuplet scalar cases, respectively. Next, we considered indirect
detection constraints, especially relevant due to potentially large
Sommerfeld enhancements. In fact, we found that the annihilation cross
sections for the $W^+W^-$ and $\gamma\gamma$ channels are considerably
affected by the Sommerfeld effect.  The quintuplet fermion DM can
evade the strong constraints of the gamma-ray experiments at around
only $m_{\chi^0}\approx2~\mathrm{TeV}$ and $7.5~\mathrm{TeV}$ due to
the drastic decrease of the cross sections by the Ramsauer-Townsend
effect.  For the septuplet scalar DM, the same thing occurs and the DM
is predicted to be $m_{\chi^0}\approx5.5~\mathrm{TeV}$ or
$7~\mathrm{TeV}$.
In addition, if more detailed analyses of the gamma-ray
constraints in MDM scenarios are taken into account, our scenario 
would be allowed only at specific DM masses for cored DM profiles.


\section*{Acknowledgments}
The authors would like to thank Asmaa Abada, Marco Cirelli, Renato
M. Fonseca, Koichi Hamaguchi, Filippo Sala and Marco Taoso for fruitful
discussions and Florian Staub for assistance in the implementation of
septuplets in {\tt SARAH}~\cite{Staub:2013tta}.  The work of M.~A. is
supported in part by the Japan Society for the Promotion of Sciences (JSPS)
Grant-in-Aid for Scientific Research (Grant No. 25400250 and
No. 26105509).  T.~T. acknowledges support from the European ITN
project (FP7-PEOPLE-2011-ITN, PITN-GA-2011-289442-INVISIBLES) and P2IO
Excellence Laboratory.  Numerical computation in this work was carried
out at the Yukawa Institute Computer Facility.
This research was partially supported by the Munich Institute for Astro-
and Particle Physics (MIAPP) of the DFG cluster of excellence ``Origin
and Structure of the Universe''.



\appendix

\section{$SU(2)$ Multiplet Notation}
\label{sec:app}

There are two common ways to denote $SU(2)$ multiplets:

\begin{itemize}
\item {\bf Tensor notation:} This is the usual choice, see
  Refs.~\cite{Hisano:2013sn,Alvarado:2014jva}. A multiplet $\Phi$ would
  be represented by a symmetric tensor with some indices,
  (four indices for quintuplet $\Phi^{ijkl}$ and six indices for septuplet $\Phi^{ijklmn}$), where all
      indices can take values $1$ or $2$. 
\item {\bf Vector notation:} This is the choice made in this paper. In
  this case, the quintuplet and septuplet are simply represented by a
      vector of $5$ and $7$ elements.
\end{itemize}

Since both notations are equally correct, this choice is just a matter
of taste. In fact, a \emph{dictionary} that translates the analytical
expressions among notations can be easily found. Regarding the
elements of the quintuplet and septuplet, the relation between the two notations is
given by
\begin{equation}
\Phi^{ijkl}\equiv i\left(
\begin{array}{c}
+\Phi^{1111}\\
+2\Phi^{1112}\\
-\sqrt{6}\Phi^{1122}\\
+2\Phi^{1222}\\
+\Phi^{2222}
\end{array}
\right),\qquad
\Phi^{ijklmn}\equiv i\left(
\begin{array}{c}
+\Phi^{111111}\\
+\sqrt{6}\Phi^{111112}\\
+\sqrt{15}\Phi^{111122}\\
-\sqrt{20}\Phi^{111222}\\
-\sqrt{15}\Phi^{112222}\\
+\sqrt{6}\Phi^{122222}\\
-\Phi^{222222}
\end{array}
\right).
\end{equation}
This allows us to write the quintuplet and septuplet $\chi$ as shown in
Eq. \eqref{eq:chi}.

We now comment on $SU(2)$ septuplet direct products. The product
${\bf7}\otimes{\bf7}$ can be decomposed as
\begin{equation}
{\bf7}\otimes{\bf7}=
{\bf13}_S\oplus{\bf11}_A\oplus{\bf9}_S\oplus{\bf7}_A
\oplus{\bf5}_S\oplus{\bf3}_A\oplus{\bf1}_S,
\end{equation}
where the indices $S$ and $A$ mean symmetric and anti-symmetric
contractions.
The symmetry properties of septuplet contractions are fundamental in
order to determine the number of relevant scalar couplings. For
example, when one considers $\chi^4$, four kinds of singlets are
obtained since anti-symmetric parts vanish:\footnote{In principle,
  when more than one singlet can be obtained for a specific Lagrangian
  term, one must check whether they are linearly independent in order
  to avoid the introduction of redundant couplings.}
\begin{eqnarray}
&&\hspace{-0.2cm}
{\bf7}\otimes{\bf7}\otimes{\bf7}\otimes{\bf7}\nonumber\\
&=&\hspace{-0.2cm}
\left({\bf13}\oplus{\bf9}\oplus{\bf5}\oplus{\bf1}\right)\otimes
\left({\bf13}\oplus{\bf9}\oplus{\bf5}\oplus{\bf1}\right)\nonumber\\
&\supset&\hspace{-0.2cm}
{\bf1}\oplus{\bf1}'\oplus{\bf1}''\oplus{\bf1}'''.
\end{eqnarray}
However one can check that only two of them are linearly independent. 
%


\end{document}